\documentclass[aps,prl,twocolumn,tightenlines]{revtex4}
\usepackage{graphicx}% Include figure files
\usepackage{dcolumn}% Align table columns on decimal point
\usepackage{bm}% bold math
\usepackage{epsfig}
\usepackage{amsmath}
\input{epsf}
\usepackage{psfrag}
\usepackage{xcolor}
\usepackage{grffile}
\usepackage{graphicx}
\usepackage{multirow}
\usepackage{bm}
\usepackage{bbm}
\usepackage{color}
\usepackage{slashed}

\newcommand{\bea}{\begin{eqnarray}}
\newcommand{\eea}{\end{eqnarray}}

\begin{document}

%\title{Hunting for Low-lying Vector Beauty-charm Meson via Polarization Information}
%\title{Distinguishing Low-lying Vector Beauty-charm Meson via Polarization Information}
\title{Distinguishing Low-lying Vector Beauty-charm Meson via Polarization Analysis}
\author{  Yiqi Geng$^{1}$, Mingqi Cao$^{1}$, Ruilin Zhu$^{1,2,3}$\footnote{Corresponding author: rlzhu@njnu.edu.cn}}
\affiliation{
$^1$ Department of Physics and Institute of Theoretical Physics, Nanjing Normal University, Nanjing, Jiangsu 210023, China\\
$^2$ CAS Key Laboratory of Theoretical Physics, Institute of Theoretical Physics,
Chinese Academy of Sciences, Beijing 100190, China\\
$^3$ Peng Huanwu Innovation Research Center, Institute of Theoretical Physics,
Chinese Academy of Sciences, Beijing 100190, China}

\begin{abstract}
To distinguish the low-lying vector beauty-charm meson, we systematically study the $B_c^*\to B_c+\gamma$, $B_c^*\to \ell+{\nu}_{\ell}$ and $B_c^{(*)}\to J/\psi+nh$ processes within effective theory by the helicity decomposition method. The significant difference of polarization asymmetry in $B_c^{(*)}\to J/\psi+nh$ indicates a general law in vector-to-vector and pseudoscalar-to-vector transition processes, which can be tested in current and future LHC experiments. In the end, we discuss the experiment search and discovery potential for the low-lying vector beauty-charm meson.

\end{abstract}

\maketitle

\textbf{\emph{Introduction.}}
Understanding of Quantum Chromodynamics (QCD) color confinement is one of the fundamental goals of
particle physics. The low-lying vector beauty-charm meson $B_c^*$ is believed to exist in various quark models and lattice simulations
of first principles QCD,
however, which has not been identified in particle experiments.  The $B_c^*$ meson has become the last missing piece of the low-lying vector meson
spectroscopy puzzle
since the next to last vector ground state $B_s^*$ was probed by CUSB-II detector in 1990~\cite{Lee-Franzini:1990dev}.

The long-standing difficulties to discover the $B_c^*$ meson come from two aspects.
On the one hand, the $B_c^*$ meson is produced in large quantities at hadron colliders
while observation of the major decay channel $B_c^*\to B_c+\gamma$ is extremely difficult due to
the low energy of the photon. The complete determination of both the emitted photon energy
and the decay width is not given in literatures. Recently, the second and third members of beauty-charm meson
family, i.e. first radially excited pseudoscalar and vector states  $B_c(2S)$ and $B_c^*(2S)$, have just discovered and confirmed
by the investigating the $B_c+2\pi$ invariant mass spectrum in CMS~\cite{CMS:2019uhm} and LHCb~\cite{LHCb:2019bem} experiments
after previous pioneering observation of one excited peak at ATLAS detector~\cite{ATLAS:2014lga}. In both CMS and LHCb experiments,
two excited structures are observed but reconstruction of the $B_c^*(2S)$ state relies on the unknown photon due to $B_c^*(2S)\to B_c^*(\to B_c+\gamma)+2\pi$, where
the absolute mass satisfies $m_{B_c^*(2S)}=m_{B_c^*(2S)}|_{rec.}+E_\gamma$  with the missing photon energy  $E_\gamma=\Delta M_{b\bar{c}(1S)}=m_{B_c^*}-m_{B_c}$.
Thus the probe of $B_c^*$ meson will affect the final determination of $B_c^*(2S)$ absolute mass. The precise study of hyperfine mass splitting is also helpful
to understand the low energy effective theory of QCD.

On the other hand, the partial decay widths of weak decay channels such as $B_c^*\to J/\psi+X_{H,L}$ are expected to have same order of magnitude compared to that of the ground beauty-charm meson decays $B_c\to J/\psi+X_{H,L}$ with H(L) denoting hadrons(leptons). But the weak decay rates of  $B_c^*$ is suppressed by a factor $\Gamma(B_c)/\Gamma(B_c^*)$ with magnitude around
$10^{-4}$ to $10^{-5}$.
Using the data sample corresponding to an integrated luminosity of $9fb^{-1}$, the LHCb collaboration have successfully measured 36463 $B_c\to J/\psi+X_H$ weak decay events~\cite{LHCb:2020ayi}. Thus one can expect several $B_c^*$ weak decay events in LHCb Run-2 existing data samples. However, the reconstruction of $B_c^*$ weak decay events is still challenging because the small hyperfine mass splitting of beauty-charm mesons leads to two relatively close peaks and  one peak is very high due to a large number of $B_c$  decay events.

In this Letter, we present the important polarization analysis of $B_c^{*}$ electromagnetic and weak decays.
We generalize the low energy effective theory for heavy quarkonium electromagnetic interactions into unequal quark mass case. The decay width of radiative decay $B_c^*\to B_c+\gamma$ is investigated in a model-independent way, where the dependence of
 $B_c^*$ electromagnetic decay widths on the emitted photon energy is given. The weak decays of $B_c^{(*)}\to J/\psi+n\pi$
 are studied in QCD effective theory. By fitting the $B_c\to J/\psi+3\pi$ partial distribution data in LHCb experiment, we can extract the spectra function of three Pions.
We find that the spectra function of three Pions is major from two resonance contributions, i.e. the $a_1(1260)$ and $\pi_2(2005)$ states. By employing the helicity decomposition method, we find that the two kinds of channels $B_c^*\to J/\psi+n\pi$ and $B_c\to J/\psi+n\pi$ have extremely different polarization behaviors dependence of the invariant
mass of $n\pi$ system.  The  $B_c^*$ meson can be distinguished in $J/\psi+n\pi$ invariant
mass distributions by introducing  a new polarization observable and measuring its value in particle experiments at LHC.

\textbf{\emph{Radiative decay.}}
The lifetime of vector beauty-charm meson $B_c^*$ is greatly shorter than that of the ground pseudoscalar beauty-charm meson $B_c$,
since $B_c^*$ meson can first radiate into $B_c$ meson with several tens of $MeV$ phase space while $B_c$ meson has to weak decay.

For the transition of doubly heavy quark mesons, potential nonrelativistic QCD (pNRQCD) is a powerful model-independent effective theory. Its lagrangian can be obtained by integrating out quarks and gluons of momentum and energy at
 order of heavy quark mass and heavy quark relative momentum from QCD~\cite{Brambilla:1999xf}. In pNRQCD effective theory, two fields $S=S(r,R,t)$ and $O=O(r,R,t)$ denoting the color singlet and octet
 quark-antiquark states respectively
 are introduced. $r$ is the heavy quark relative coordinate while $R$ is the center of mass coordinate.
In equal quark mass case, the  pNRQCD Lagrangian relevant
 to describe the magnetic dipole transition at order $E^3_\gamma v^2/m^2$ is systematically established in Ref.~\cite{Brambilla:2005zw}, where
 the radiative decay width is obtained as $\Gamma_{J/\psi \rightarrow \eta_c +\gamma}=(1.5\pm1.0)keV$ in excellent agreement with experimental data. We generalize
 the pNRQCD Lagrangian into unequal quark mass case, and then the effective Lagrangian at order $E^3_\gamma v^2/m^2$  can be written as
\begin{align}
& \mathcal{L}_{\gamma \mathrm{pNRQCD}}=\int d^3 r \operatorname{Tr}\left[e \frac{e_Q -e_Q'}{2}V_A^{\mathrm{em}} \mathrm{S}^{\dagger} \mathbf{r} \cdot \mathbf{E}^{\mathrm{em}} \mathrm{S}\right.\nonumber \\
& +e(\frac{ e_Q m_Q'-e_Q' m_Q}{4 m_Q m_Q'}) \left[V_S^{\frac{\sigma \cdot B}{m}}\left\{\mathrm{~S}^{\dagger}, \boldsymbol{\sigma} \cdot  \mathbf{B}^{\mathrm{em}}\right\} \mathrm{S} \right. \nonumber\\
& +\frac{1}{8} V_S^{(r \cdot \nabla)^2 \frac{\sigma \cdot B}{m}}\left\{\mathrm{~S}^{\dagger}, \mathbf{r}^i \mathbf{r}^j\left(\boldsymbol{\nabla}^i \nabla^j \boldsymbol{\sigma} \cdot \mathbf{B}^{\mathrm{em}}\right)\right\} \mathrm{S} \nonumber\\
& \left.+ V_O^{\frac{\sigma \cdot B}{m}}\left\{\mathrm{O}^{\dagger}, \boldsymbol{\sigma} \cdot  \mathbf{B}^{\mathrm{em}}\right\} \mathrm{O}\right]\nonumber \\
& +e(\frac{ e_Q m^2_{Q'}-e_Q' m^2_Q}{32 m_Q^2 m^2_{Q'}}) \left[4\frac{V_S^{\frac{\sigma \cdot B}{m^2}}}{r}\left\{\mathrm{~S}^{\dagger}, \boldsymbol{\sigma} \cdot \mathbf{B}^{\mathrm{em}}\right\} \mathrm{S} \right.\nonumber\\
& +4\frac{V_S^{\frac{\sigma \cdot(r \times r \times B)}{m^2}}}{r}\left\{\mathrm{~S}^{\dagger}, \boldsymbol{\sigma} \cdot\left[\hat{\mathbf{r}} \times\left(\hat{\mathbf{r}} \times \mathbf{B}^{\mathrm{em}}\right)\right]\right\} \mathrm{S} \nonumber\\
& - V_S^{\frac{\sigma \cdot \nabla \times E}{m^2}}\left[\mathrm{~S}^{\dagger}, \boldsymbol{\sigma} \cdot\left[-i \boldsymbol{\nabla} \times, \mathbf{E}^{\mathrm{em}}\right]\right] \mathrm{S} \nonumber\\
& \left. - V_S^{\frac{\sigma \cdot \nabla_r \times r \cdot \nabla E}{m^2}}\left[\mathrm{~S}^{\dagger}, \boldsymbol{\sigma} \cdot\left[-i \boldsymbol{\nabla}_r \times, \mathbf{r}^i\left(\boldsymbol{\nabla}^i \mathbf{E}^{\mathrm{em}}\right)\right]\right] \mathrm{S} \right]\nonumber\\
& +e(\frac{ e_Q m^3_{Q'}-e_Q' m^3_Q}{8 m_Q^3 m^3_{Q'}})\left[V_S^{\frac{\nabla_r^2 \sigma \cdot B}{m^3}}\left\{\mathrm{~S}^{\dagger}, \boldsymbol{\sigma} \cdot \mathbf{B}^{\mathrm{em}}\right\} \nabla_r^2 \mathrm{~S}\right.\nonumber \\
& \left.\left.+ V_S^{\frac{(\nabla r \cdot \sigma)(\nabla r \cdot B)}{m^3}}\left\{\mathrm{~S}^{\dagger}, \boldsymbol{\sigma}^i\mathbf{B}^{\mathrm{em} j}\right\} \boldsymbol{\nabla}_r^i \nabla_r^j \mathrm{~S}\right]\right],
\end{align}
where $Q$ and $Q'$ denote two different heavy quarks.

The final result for the decay width of $B_c^*(p)\to B_c(p')+\gamma(k)$ is
\begin{align}
 \Gamma_{B_c^* \rightarrow B_c +\gamma}&=\frac{ \alpha(e_Q m_Q'-e_Q' m_Q)^2E^3_\gamma}{3 m^2_Q m^2_{Q'}}V_S^{\frac{\sigma \cdot B}{m}}\left(1-\frac{E_\gamma}{m_{B_c^*}}\right),
\end{align}
where  the photon energy is expressed as $E_\gamma=\frac{m_{B_c^*}^2-m_{B_c}^2}{2m_{B_c^*}}$. The matching coefficient is known at one loop with $V_S^{\frac{\sigma \cdot B}{m}}=1+C_F\frac{\alpha_s}{2\pi}$~\cite{Manohar:1997qy}.
Other higher-order pNRQCD operators are not considered here, however, one expect that their contributions are small in similar to the case in bottomonium. We can choose $Q=b$ and $Q'=c$ for beauty and
charm quarks in the following.

\begin{figure}[th]
\includegraphics[width=0.45\textwidth]{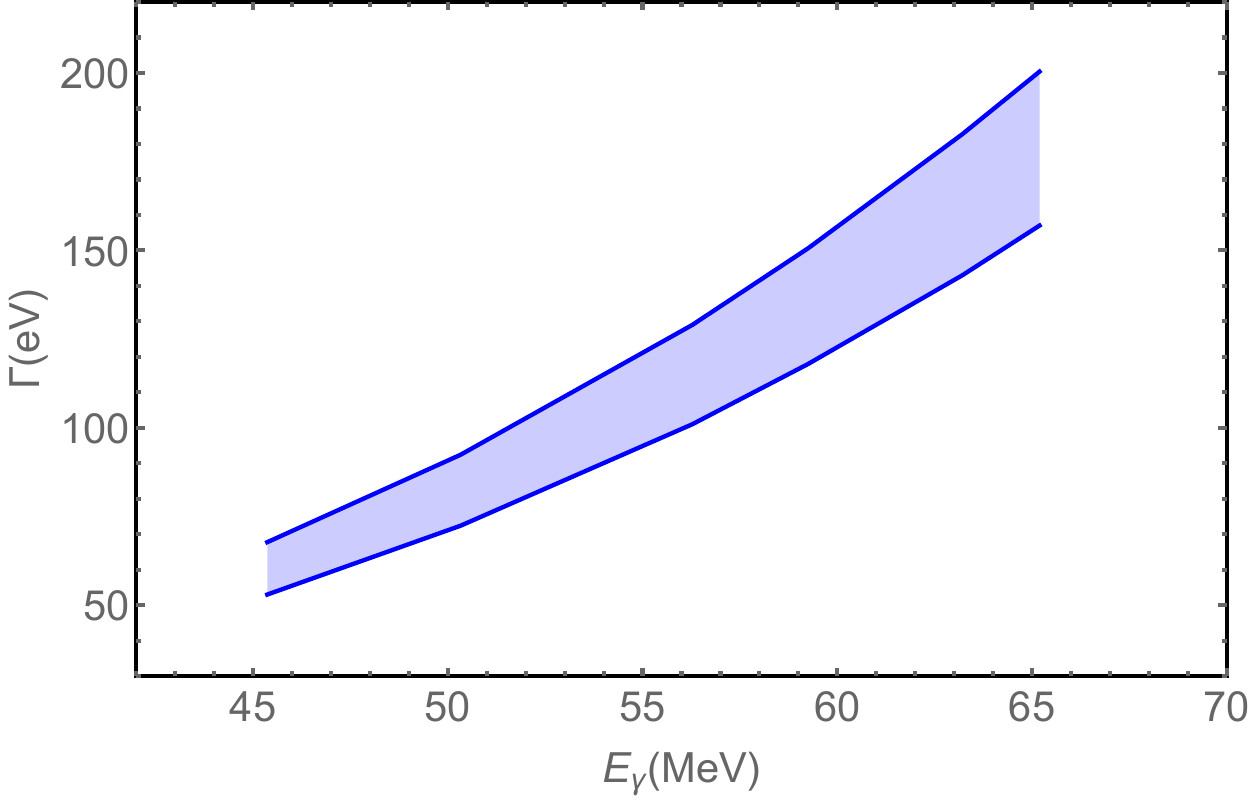}
\caption{ Total decay width of the low-lying vector $B_c^*$ meson as functions of the emitted photon energy.}\label{fig:decayw}
\end{figure}

The total decay width of the vector $B_c^*$ meson can be approximated as $\Gamma\simeq \Gamma_{B_c^* \rightarrow B_c +\gamma}$ since other weak decay channels have a suppression factor $\Gamma(B_c)/\Gamma(B_c^*)$ with magnitude around
$10^{-4}$ to $10^{-5}$. Consider that the heavy quark pole mass is usually chosen as $m_b=4.8\pm 0.2 GeV$ and $m_c=1.6\pm0.1GeV$, the total decay width of the vector $B_c^*$ meson as functions of the emitted photon mass or hyperfine mass splitting is plotted in Fig.~\ref{fig:decayw}.
If we choose the vector $B_c^*$ meson mass as $6331(4)(6)MeV$ from Lattice QCD simulation~\cite{Mathur:2018epb}, the total decay width of the vector $B_c^*$ meson is estimated as $\Gamma=114^{+60}_{-42} eV$ where the large uncertainty is from the sensitivity of decay width on meson mass. One should note that there are already several theoretical predictions in literatures~\cite{Ebert:2002pp,Fulcher:1998ka,Gershtein:1994jw,Godfrey:2004ya,Eichten:1994gt}, however the model-independent investigation is first given in our paper.

In calculation, only two polarization states of the vector $B_c^*$ meson with $|J=1, \lambda=\pm 1\rangle$ are equally contributed in the radiative $B_c^*$ meson decays. In the rest frame of final $B_c$ meson, the angular momentum projection is identical to the vector $B_c^*$ meson helicity $\lambda$. This phenomenon
can be understood by the conservation of angular momentum and parity. Since the photon only has two transversally polarization statues,
the initial $B_c^*$ meson with $|J=1, \lambda=0\rangle$ can not emit a photon parallel to momentum direction and thus is forbidden in its radiative decay. The right-hand circularly polarized photon is emitted when the initial $B_c^*$ meson with $|J=1, \lambda=1\rangle$ decays into zero-spin $B_c$ meson.
While the left-hand circularly polarized photon is emitted when the initial $B_c^*$ meson with $|J=1, \lambda=-1\rangle$ decays into zero-spin $B_c$ meson.

\textbf{\emph{Weak Decay.}}
In the radiative decay of $B_c^*$ meson with $B_c^* \rightarrow B_c +\gamma$,
the transverse polarization of $B_c^*$ meson with $|J=1, \lambda=\pm 1\rangle$ contributes, while the longitudinal polarization of $B_c^*$ meson with $|J=1, \lambda=0\rangle$ decouples.
In the weak decays, both the transversely and longitudinally polarized $B_c^*$ mesons will come in and contribute the Feynman amplitudes.

The pure leptonic
 weak decays  $B^*_c\to \ell+{\nu}_{\ell}$ have been studied up to three-loop accuracy in  Refs.~\cite{Tao:2022qxa,Tao:2023mtw}, where the branching ratios are given with magnitude around $10^{-6}$. If we focus on the polarization decomposition, the transverse and longitudinal  polarizations of $B_c^*$ meson leptonic decay widths are
 \begin{align}
	\Gamma(B_c^{*}(\lambda=\pm1) \to \ell {\nu}_{\ell})=&\frac{{|V_{cb}|}^2}{12\pi}  G_F^2 {f_{B_c^*}^2} \left(1-\frac{m_\ell^2}{m_{B_c^*}^2}\right)^2
\nonumber\\&\times m_{B_c^*}^3 ,\nonumber\\
	\Gamma(B_c^{*}(\lambda=0) \to \ell  {\nu}_{\ell})=&\frac{m_\ell^2\Gamma(B_c^{*+}(\lambda=\pm1) \to \ell {\nu}_{\ell})}{2m_{B_c^*}^2} ,
\end{align}
where the factor $(m_\ell/m_{B_c^*})^2$ in longitudinal polarization formula represents a helicity suppression which
is just  a consequence of angular momentum conservation.

The weak decays into $B^*_c\to J/\psi+X_{H,L}$ are also good channels to probe the $B_c^*$  meson at hadron colliders.
The vector current form factors of $B_c^*$ into $J/\psi$ can be defined as
\begin{align}
&\left\langle J/\psi\left(\epsilon^{\prime }, p^{\prime }\right)\left|\bar{b} \gamma_\mu c\right| B_c^*\left(\epsilon, p\right)\right\rangle\nonumber\\
= & -\left(\epsilon \cdot \epsilon^{ \prime *}\right)\left[P_\mu V_1\left(q^2\right)-q_\mu V_2\left(q^2\right)\right] -\left(\epsilon\cdot q\right) \epsilon_\mu^{ \prime *} V_3\left(q^2\right)\nonumber\\&+\left(\epsilon^{\prime *} \cdot q\right) \epsilon_\mu V_4\left(q^2\right)+\frac{\left(\epsilon \cdot q\right)\left(\epsilon^{ \prime *} \cdot q\right)}{M^{ 2}-M^{\prime 2}}\left[\left(P^\mu \right.\right.\nonumber\\&\left.\left.-\frac{M^{ 2}-M^{\prime 2}}{q^2} q^\mu\right) V_5\left(q^2\right)+\frac{M^{ 2}-M^{ \prime 2}}{q^2} q^\mu V_6\left(q^2\right)\right],
\end{align}
where $P=p+p^{\prime }$, $q=p-p^{\prime }$. $M$ and $M'$ are the masses of $ B_c^*$ and $J/\psi$ respectively.
 Similarly, the axial-vector current form factors of $B_c^*$  into $J/\psi$ can be defined as
\begin{align}
&\left\langle J/\psi\left(\epsilon^{\prime }, p^{\prime }\right)\left|\bar{b} \gamma_\mu \gamma_5 c\right| B_c^*\left(\epsilon, p\right)\right\rangle\nonumber\\= & i \varepsilon_{\mu \nu \alpha \beta} \epsilon^{\alpha} \epsilon^{ \prime * \beta}\left[P^\nu A_1\left(q^2\right)+ q^\nu A_2\left(q^2\right)\right] \nonumber\\
& +\frac{i \varepsilon_{\mu \nu \alpha \beta} P^\alpha q^\beta}{M^{ 2}-M^{ \prime 2}}\left[\epsilon^{\prime *} \cdot q \epsilon^{ \nu} A_3\left(q^2\right)-\epsilon \cdot q \epsilon^{ \prime * \nu} A_4\left(q^2\right)\right].
\end{align}

The differential distribution for $B^{(*)}_c\to J/\psi+nh$ can be decomposed into
\begin{align}
\frac{d\Gamma(B^{(*)}_c\to J/\psi+nh)}{dq^2}= &\sum_{\lambda_i} \frac{{|V_{cb}|}^2 G_F^2 a_1^2 |\mathbf{p'}|}{32 \pi  M^2}\Gamma_{J_1\lambda_1J_2\lambda_2\lambda_{nh}} ,\label{helicity-formula}
\end{align}
where the parameter $\Gamma_{J_1\lambda_1J_2\lambda_2\lambda_{nh}}$ is the helicity component with the initial meson angular momentum $J_1$ and the $J/\psi$ angular momentum $J_2$. The $J/\psi$ moving momentum is $|\mathbf{p'}|=((M^2+M'^2-q2)^2/(4M^2)- M'^2)^{1/2}$.  Due to the angular momentum conservation, we have the following nontrivial helicity components
\begin{align}
\Gamma_{11110}= &2 \left[V_1^2 \left(\left(M-M'\right)^2-q^2\right)
   \left(\left(M'+M\right)^2-q^2\right)\right.\nonumber\\&\left.+\left(A_1 \left(M^2-M'^{2}\right)+A_2
   q^2\right){}^2\right]\rho^{nh}_T(q^2),\\
\Gamma_{1111t}= &2 \left[A_1^2 \left(-2 M^2 \left(M'^{2
   }+q^2\right)+\left(M'^{2 }-q^2\right)^2+M^4\right)\right.\nonumber\\&\left.+\left(V_1 \left(M'^{2 }-M^2
   \right)+q^2 V_2\right){}^2\right]\rho^{nh}_L(q^2) ,
\end{align}
other four nontrivial helicity components  for vector $B_c^*$ decay are complicated, which are given in Appendix. Due to symmetry, $\lambda_{1,2}=1$ represents
$\lambda_{1,2}=\pm1$. For pseudoscalar $B_c$ decay,
there are similar helicity components
\begin{align}
\Gamma_{00100}= &
\frac{\rho^{nh}_T(q^2)}{4 M'^2
   \left(M'+M\right){}^2}\left[-A'_2 \left(M^4-2 M^2
   \left(M'^2+q^2\right)\right.\right.\nonumber\\&\left.+\left(M'^2-q^2\right){}^2\right)
   \nonumber\\&\left.+A'_1 \left(M'+M\right){}^2 \left(M^2-M'^2-q^2\right)\right]{}^2,\\
\Gamma_{0010t}= &
\rho^{nh}_L(q^2){A'}_0^2 \left[-2 M^2
   \left(M'^2+q^2\right)+M^4\right.\nonumber\\&\left.+\left(M'^2-q^2\right){}^2\right],\\
\Gamma_{00111}= &
\frac{2q^2 \rho^{nh}_T(q^2)}{\left(M'+M\right){}^2} \left[ {A'}_1^2 \left(M'+M\right){}^4 +V'^2 \left(M^4
\right.\right.\nonumber\\&\left.\left.-2 M^2
   \left(M'^2+q^2\right)+\left(M'^2-q^2\right){}^2\right)\right],
\end{align}
where the definition of $B_c\to J/\psi$ form factors $A'_i(q^2)$ and $V'(q^2)$ can be found in Eqs. (2-3) in Ref.~\cite{Wang:2018duy}.

The above spectral functions $\rho^{nh}_{T,L}(q^2)$ are universal and can be defined as
\begin{align}
\int \frac{d \Phi(W^*\to nh)}{2\pi}\epsilon^{nh}_\mu \epsilon^{*nh}_\nu= &\left(q_\mu q_\nu-q^2g_{\mu\nu}\right)\rho^{nh}_{T}(q^2)\nonumber\\
&+q_\mu q_\nu\rho^{nh}_{L}(q^2) .
\end{align}
In principle, the dimensionless spectral functions $\rho^{nh}_{T,L}(q^2)$ can be determined from nonperturbative calculation or experimental data.
The LHCb collaboration have studied the $B_c^+\to J/\psi+\pi^++\pi^-+\pi^+$ process and measured the $3\pi$ invariant mass distribution in Ref.~\cite{LHCb:2021tdf}. The polarization measurement of $J/\psi$ is not performed in this process, however, the $3\pi$ distribution has
a large peak around $a_1(1260)$ and a small peak around $\pi_2(2005)$ in Fig.~2. Thus we can conclude that the spectral function $\rho^{nh}_{T}(q^2)$
dominates in $B_c\to J/\psi+3\pi$.
 \begin{figure}[th]
        \includegraphics[width=0.4\textwidth]{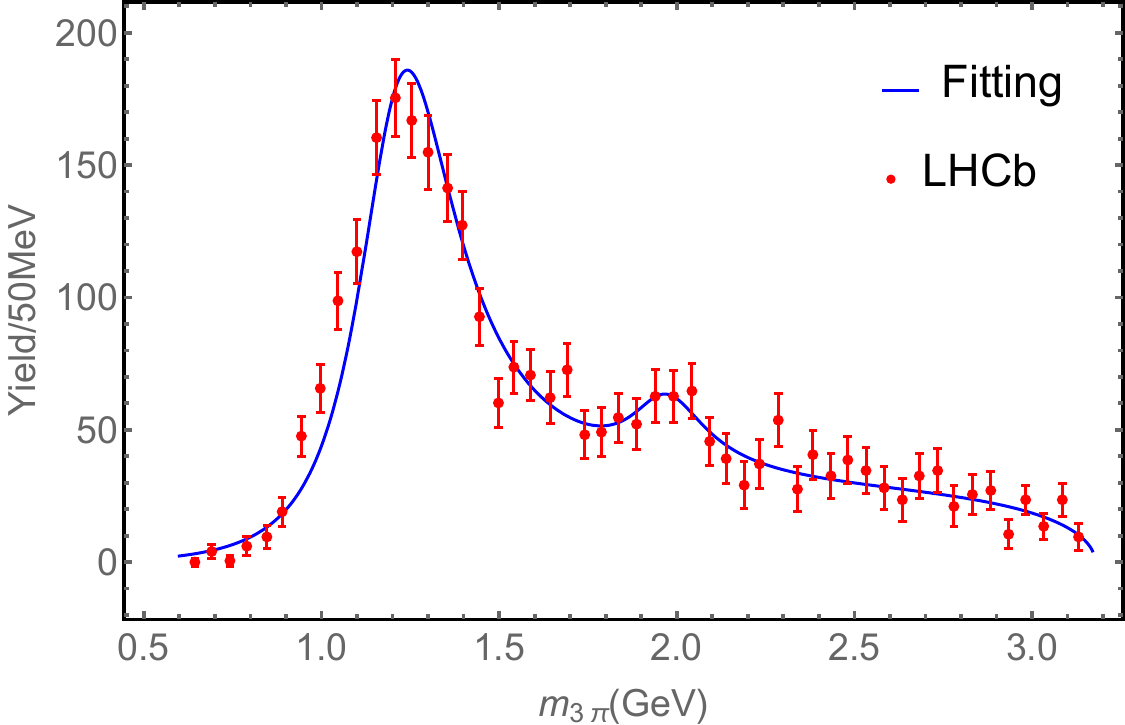}
        \caption{The invariant mass distribution of $m_{3\pi}$ in the $B_c^+\to J/\psi\pi^+\pi^-\pi^+$ process. The red ones are from the LHCb measurements based on the sample corresponding to an integrated luminosity of $9fb^{-1}$ data~\cite{LHCb:2021tdf}, while the blue line is our fitting result.}
        \label{fitting}
    \end{figure}

We use the following parametrization form for  $\rho_T^{3\pi}$
   \begin{align}
             \rho_T^{3\pi}(m^2)=&\frac{a}{2m} (\frac{m^2 - m^2_{\pi}e}{m^2})^{-2}(1 -   m^2f)
             \nonumber\\& \times[ \frac{1}{b^2/4 + (m - m_{1})^2} +\frac{c}{d^2/4 + (m - m_{2})^2} ],
 \end{align}
 which is different to the form in Ref.~\cite{Luchinsky:2012rk,Luchinsky:2022pxu}.
If we input the two poles $m_{1}=1.209GeV$ and $m_{2}=1.963GeV$ for $a_1(1260)$ and $\pi_2(2005)$ peaks, the chi-square goodness of fitting is $\chi^2/dof=1.65$.
The parameters are fitted as $a=0.12GeV$, $b=0.341GeV$, $c=0.021$, $d=0.256GeV$, $e=-12.456$, and $f=-0.069GeV^{-2}$.  Therein $b=0.341GeV$ and $d=0.256GeV$
can explain the decay width of $a_1(1260)$ and $\pi_2(2005)$  respectively. For future theoretical and experimental studies,  this process can be employed to precisely measure the basic
quantities for both the $a_1(1260)$ and $\pi_2(2005)$ states. Note that the value $a=0.12GeV$ is obtained by considering
the $B_c$ hadroproduction cross section around $100nb^{-1}$ at LHC in Ref.~\cite{Chang:2003cr}.
\begin{figure}[th]
        \includegraphics[width=0.4\textwidth]{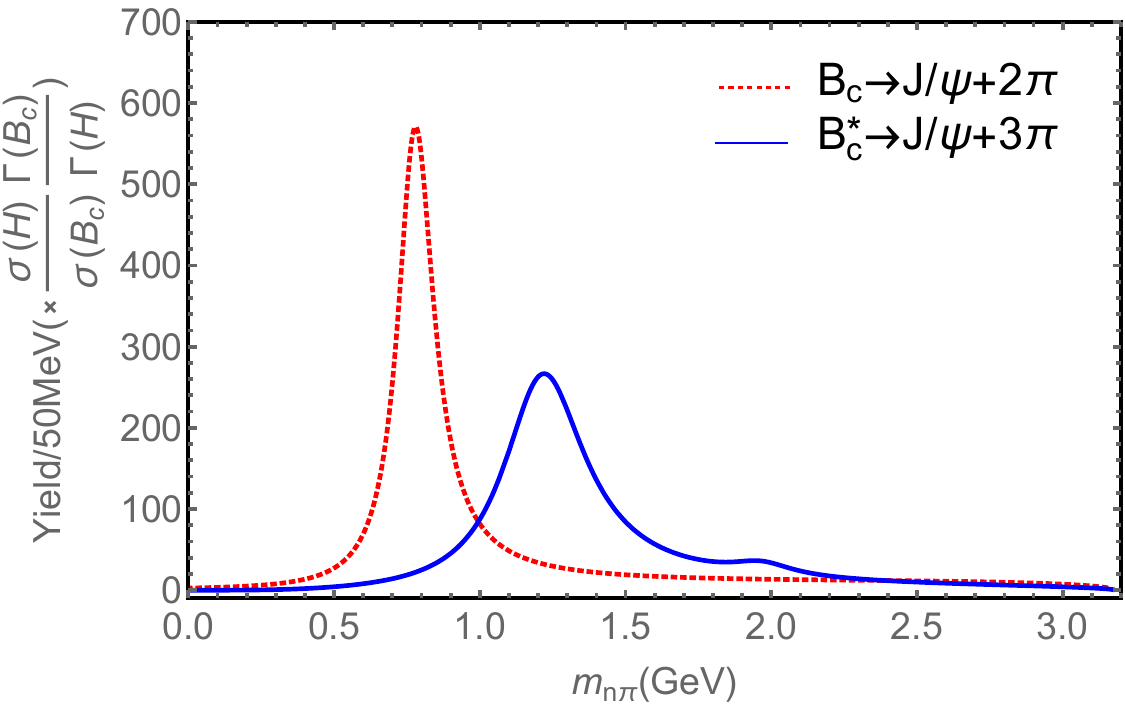}
        \caption{The reconstruction events distribution for both the $B_c\to J/\psi+2\pi$ and $B_c^*\to J/\psi+3\pi$ at LHCb run-2.}
    \end{figure}

 \begin{figure}[th]
        \includegraphics[width=0.4\textwidth]{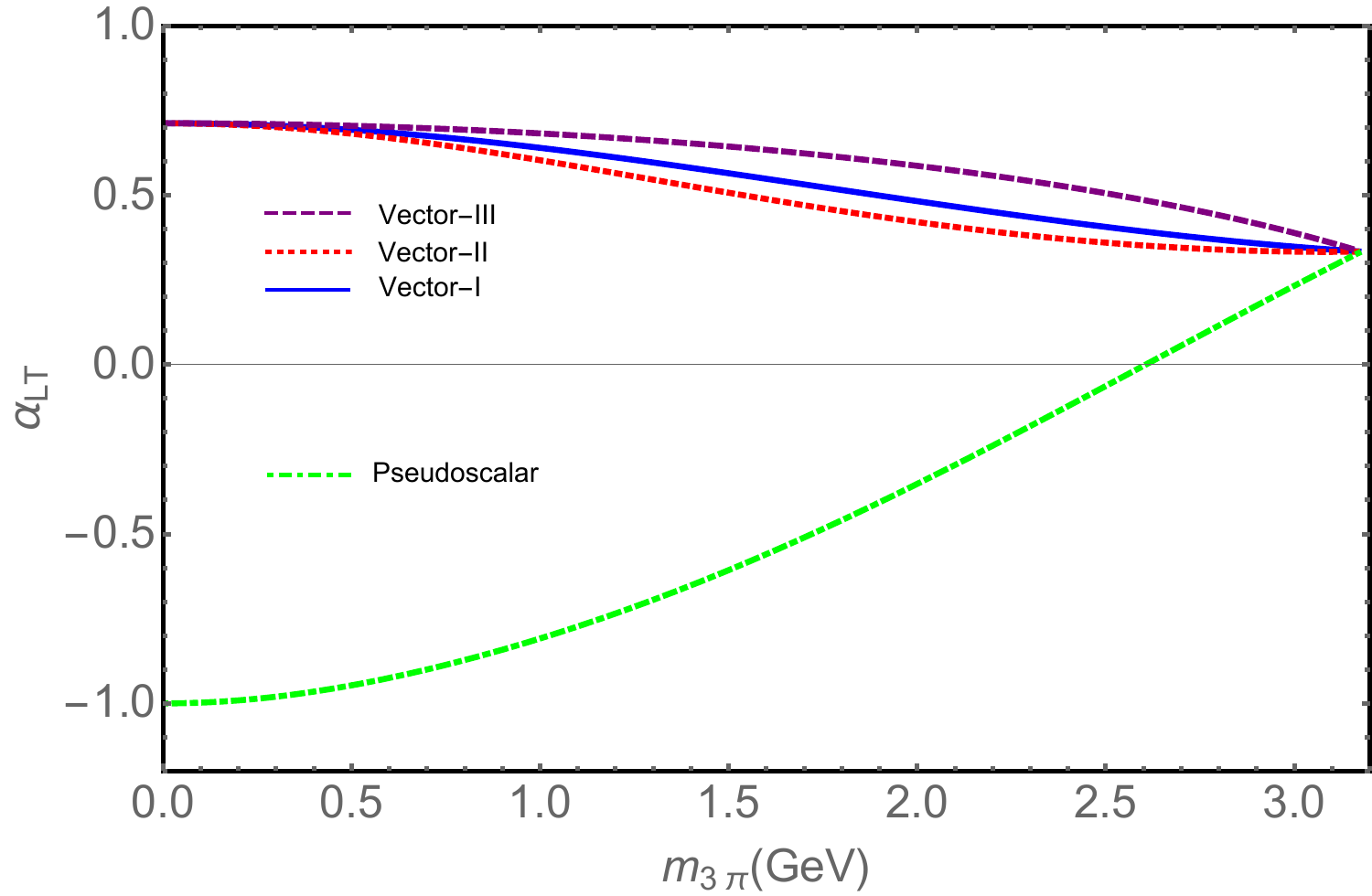}
        \caption{The polarization asymmetry $\alpha_{LT}$ significantly different in the two processes $B_c^*\to J/\psi+n\pi$ and $B_c\to J/\psi+n\pi$ with
        $n=2$ or $n=3$.  In $B_c\to J/\psi+n\pi$  decays, the  polarization asymmetry $\alpha_{LT}$ denoting ``Pseudoscalar'' do not sensitive to the relative magnitude between the $n\pi$
        spectra functions $\rho_{L,T}(q^2)$.  While in $B^*_c\to J/\psi+n\pi$  decays, the  polarization asymmetry $\alpha_{LT}$  is slightly sensitive to the
        spectra functions, where ``Vector-I, Vector-II, Vector-III'' represent $\rho_{L}=\rho_{T}$, $\rho_{L}=\rho_{T}/10$, $\rho_{L}=10\rho_{T}$, respectively. }
    \end{figure}

Employing the $\rho(770)$ dominant model for  $\rho_T^{2\pi}$
   \begin{align}
             \rho_T^{2\pi}(m^2)=&\frac{a'}{\Gamma_\rho^2/4 + (m - m_{\rho})^2} ,
 \end{align}
where the parameter $a'=0.1198GeV^{2}$ can be extracted from the theoretical prediction of $B_c\to J/\psi+\rho$ in Ref.~\cite{Qiao:2012hp}.

The nontrivial results of various form factors at leading-order can be calculated in NRQCD~\cite{Bodwin:1994jh}
\begin{align}
&V_1(y)=\frac{128 \pi  (z+1)^{5/2} \alpha _s\phi_{^1 S_0[c\bar{c}]}^{(0)}(0)\phi_{^1 S_0[c\bar{b}]}^{(0)}(0)}{3 z^{3/2} m_b^3 (y-z+1)^2 (y+z-1)^2},\nonumber\\
&V_3(y)=2V_1(y)=2A_1(y),\nonumber\\
&V_2(y)=A_2(y)=\frac{1-z}{1+z}V_1(y),\nonumber\\
&V_4(y)=\frac{4z}{1+z}V_1(y),
\end{align}
where $z=m_c/m_b$ and $y=\sqrt{q^2/m_b^2}$.  The HPQCD collaboration have
performed the first lattice calculation of $B_c\to J/\psi$ form factors~\cite{Harrison:2020gvo}. In previous works~\cite{Tang:2022nqm,Shen:2021dat,Wang:2018duy,Qiao:2012hp,Qiao:2012vt}, various $B_c\to J/\psi$ form factors have been systematically studied by the NRQCD+HPQCD approach along with the BGL parametrization method~\cite{Boyd:1997kz}. Similarly, we can further determine the $B^*_c\to J/\psi$  form factors after combining the lattice QCD results of $B_c\to J/\psi$ form factors and NRQCD relations among form factors. High-order calculation affects the NRQCD relations among form factors very slightly.

According to the LHCb reconstruction efficiency in Ref.~\cite{LHCb:2021tdf}, the event yields per very 50MeV can also be obtained for both the $B_c\to J/\psi+2\pi$ and $B_c^*\to J/\psi+3\pi$ processes, which
have been plotted in Fig. 3.

The helicity formula for the decay width is given in Eq.~(\ref{helicity-formula}). One can further define the polarization asymmetry $\alpha_{LT}$ as
\begin{align}
\alpha_{LT}=&\sum_{\lambda_1,\lambda_{nh}}\frac{\Gamma_{J_1\lambda_111\lambda_{nh}}-\Gamma_{J_1\lambda_110\lambda_{nh}}}{\Gamma_{J_1\lambda_111\lambda_{nh}}+\Gamma_{J_1\lambda_110\lambda_{nh}}},
\end{align}
where we only observe the transverse  and longitudinal polarization of $J/\psi$ due to its feasibility at particle experiments.

We plot the results of $\alpha_{LT}$ of  $B_c^*\to J/\psi+n\pi$ and $B_c\to J/\psi+n\pi$ with
        $n=2$ or $n=3$ in Fig.~4. In calculation, we found that $J/\psi$ prefers to be longitudinally polarized in $B_c$ decays while
        $J/\psi$ prefers to be transversely polarized in $B^*_c$ decays, which indicates a general law  of polarization asymmetry for pseudoscalar(vector) meson to vector meson decays. In $P\to V$ transition, the final vector meson $V$ prefers to be longitudinally polarized and
         makes 100\% longitudinally polarized ($\alpha_{LT}=-1$) in maximum recoil point ($q^2=0$). In $V\to V'$ transition, the final vector meson $V'$ prefers to be transversely polarized and
       gets a large polarized rate ($0.5<\alpha_{LT}<1$) in maximum recoil point ($q^2=0$). This general law shall be also tested in various processes such as $B_s/B_s^*\to D_s^*+nh$,
       $B/B^*\to D^*+nh$, $D_s/D_s^*\to\phi+nh$ and $D/D^*\to K^*+nh$.

Going back to the identification of $B_c^*$ meson, one may expect  around 20  $B_c^*\to J/\psi+\pi$ and 280  $B_c^*\to J/\psi+\ell+{\nu}_{\ell}$ at LHCb run-2, assuming $9\times 10^8$  $B_c^*$ mesons are produced~\cite{Chang:2003cr}.
However, only 1 $B_c^*\to J/\psi+\pi$ and 11  $B_c^*\to J/\psi+\ell+{\nu}_{\ell}$ can be reconstructed if consider the LHCb efficiency in Ref.~\cite{LHCb:2020ayi}. The reconstruction events will increase into 33 times in future LHCb run-3 and run-4 experiments.
The polarization measurement of $J/\psi$ will eliminate the possible $B_c$ background to probe the $B_c^*$ meson. Apart from the channels in the paper, one can also probe the $B_c^*$ meson by $B_c^*\to B_s/B+n\pi$ with around $10^{-5}$ branching ratios.

\textbf{\emph{Conclusion.}}

In this Letter, the electromagnetic and weak decays of $B_c^{*}$ are studied in a model-independent way. We have shown that the helicity decomposition in  $B_c^*\to B_c+\gamma$, $B_c^*\to \ell+{\nu}_{\ell}$ and $B_c^*\to J/\psi+nh$ by the polarization analysis. The polarization asymmetry $\alpha_{LT}$ introduced in the paper is an important physical observable to distinguish the initial pseudoscalar $B_c$ and vector  $B_c^{*}$ states. It also reveals a general law in $P\to V$  and $V\to V'$ transition processes, which can be tested by the polarization measurements of the final vector mesons.
In the end, the long-sought vector $B_c^{*}$ meson has a good opportunity to be resolved during LHCb  Run-3 or Run-4 and future experiments such as CEPC running at $Z$ boson pole.

\begin{acknowledgments}
{\it Acknowledgement.}
 We thank the valuable discussions with Prof. Chao-Hsi Chang and Prof. Bing-Song Zou. This work is supported by NSFC under grant No.~12322503, No.~12047503, and No.~12075124,
 and by Natural Science Foundation of Jiangsu under Grant No.~BK20211267.
\end{acknowledgments}

\appendix
\begin{widetext}

\section{Appendix }
The other four helicity components in Equation (\ref{helicity-formula}) have the following expressions
\begin{align}
\Gamma_{10111}= &\frac{q^2 \rho^{nh}_T(q^2)}{2 M^2 \left(M^2-M'^2\right){}^2
  } \left[2 A_1
   \left(M^2-M'^2\right) \left(3 M^2+M'^2-q^2\right) \left(\left(A_2-2 A_4\right)
   M^2 q^2+\left(A_2+A_4\right) \left(M^2-M'^2\right){}^2\right.\right.\nonumber\\&\left.-\left(A_2+2 A_4\right) q^2 M'^2+A_4
   q^4\right)+\left(\left(A_2-2 A_4\right) M^2 q^2+\left(A_2+A_4\right)
   \left(M^2-M'^2\right){}^2-\left(A_2+2 A_4\right) q^2 M'^2+A_4 q^4\right){}^2\nonumber\\&\left.+A_1^2 \left(M^2-M'^2\right){}^2 \left(3 M^2+M'^2-q^2\right){}^2+V_3^2
   \left(M-M'\right){}^2 \left(M'+M\right){}^2 \left(\left(M-M'\right){}^2-q^2\right)
   \left(\left(M'+M\right){}^2-q^2\right)\right],\nonumber\\
\Gamma_{10100}= &\frac{\left(-2 M^2 \left(M'^2+q^2\right)+M^4+\left(M'^2-q^2\right){}^2\right)\rho^{nh}_T(q^2)}{16 M^2
   M'^2 \left(M^2-M'^2\right){}^2}
   \left[\left(M-M'\right) \left(M'+M\right) \left(M^2 \left(2 V_1-V_3+V_4\right)\right.\right.\nonumber\\&\left.\left.+\left(2
   V_1+V_3-V_4\right) M'^2+q^2 \left(-2 V_1+V_3+V_4\right)\right)+V_5
   \left(\left(M-M'\right){}^2-q^2\right) \left(\left(M'+M\right){}^2-q^2\right)\right]{}^2,\nonumber\\
 \Gamma_{1010t}= &  \frac{\rho^{nh}_L(q^2)}{16 M^2 M'^2}\left[2 M^2 \left(\left(-V_3+V_4+V_6\right) M'^2+q^2
   \left(V_1+V_2-V_3+V_4+V_6\right)\right)+M^4 \left(-\left(2
   V_1-V_3+V_4+V_6\right)\right)\right.\nonumber\\&\left.+\left(M'^2-q^2\right) \left(\left(2 V_1+V_3-V_4-V_6\right)
   M'^2+q^2 \left(2 V_2-V_3+V_4+V_6\right)\right)\right]{}^2,\nonumber\\
 \Gamma_{11101}= & \frac{q^2 \rho^{nh}_T(q^2)}{2
   M'^2\left(M^2-M'^2\right){}^2} \left[2 A_1
   \left(M^2-M'^2\right) \left(M^2+3 M'^2-q^2\right) \left(-q^2 \left(2 A_3
   \left(M^2+M'^2\right)+A_2 \left(M^2-M'^2\right)\right)+A_3 q^4\right.\right.\nonumber\\&\left.\left.+\left(A_2+A_3\right)
   \left(M^2-M'^2\right){}^2\right)+\left(A_3
   q^4-q^2 \left(2 A_3 \left(M^2+M'^2\right)+A_2
   \left(M^2-M'^2\right) \right)+\left(A_2+A_3\right) \left(M^2-M'^2\right){}^2\right){}^2\right.\nonumber\\&\left.+A_1^2 \left(M^2-M'^2\right){}^2 \left(M^2+3 M'^2-q^2\right){}^2+V_4^2 \left(M^2-M'^2\right){}^2
   \left(\left(M-M'\right){}^2-q^2\right) \left(\left(M'+M\right){}^2-q^2\right)\right].\nonumber
\end{align}

 \end{widetext}

\end{document}